\documentclass[twocolumn,amsmath,amssymb,superscriptaddress,aps,prc]{revtex4-1}

\usepackage{graphicx}
\usepackage{epsf}
\usepackage{amsmath,amssymb}
\usepackage{bm}
\usepackage{color}
\usepackage{ulem}

\begin{document}

\title{ 
Stability of $\alpha$-chain States against Break-up and Binary Disintegrations
}%

%
%
%
\author{Akihiro Tohsaki}

\affiliation{
Research Center for Nuclear Physics (RCNP), Osaka University,
10-1 Mihogaoka, Ibaraki, Osaka 567-0047, Japan
}

\author{Naoyuki Itagaki}

\affiliation{
Yukawa Institute for Theoretical Physics, Kyoto University, 606-8502 Kyoto, Japan 
}

\date{\today}

\begin{abstract}
We focus upon the raison d'$\hat{\rm e}$tre of the $\alpha$-chain states on the basis of the fully microscopic framework, where the Pauli principle among all the nucleons is fully taken into account. 
Our purpose is to find out the limiting number of $\alpha$ clusters, on which the linear $\alpha$-cluster state can stably exist. 
How many 
$\alpha$ clusters can stably make an $\alpha$-chain state?  We examine the properties of equally separated
$\alpha$ clusters on a straight line and compare its stability with that on a circle. We also confirm its stability in terms of break-up and binary disintegrations including $\alpha$-decay and fission modes. 
For the effective nucleon-nucleon interaction,
we employ the F1 force, which has finite-range three-body terms and
guarantees overall saturation properties of nuclei.
This interaction also gives
reasonable binding energy and size of the $\alpha$ particle and the $\alpha$-$\alpha$ scattering phase shift.
The result astonishes us because we can point out a possible existence of $\alpha$-chain states with vast numbers of $\alpha$ clusters.     
\end{abstract}

\maketitle


There has been a long history of the theoretical and experimental studies on the $\alpha$-chain state.  More than half century ago, Morinaga, who was an experimentalist, pointed out that the second $0^+$ state of $^{12}$C just above the three-$\alpha$ threshold energy was a possible candidate of it, 
and now the state is regarded as the $\alpha$-gas state~\cite{PhysRev.101.254,Uegaki12C}. 
Since there is no stable nucleus with the mass number of five or eight,
formation of  $^{12}$C from three $^4$He nuclei ($\alpha$ clusters) is a key process
of the nucleosynthesis. Here, the second $0^+$ state
at $E_x = 7.6542$ MeV plays a crucial role, which is 
the second excited state of $^{12}$C and located
just above the threshold energy  to decay into three $^4$He nuclei~\cite{Hoyle}.
The Tohsaki-Horiuchi-Schuck-R\"{o}pke (THSR) wave function has been 
widely used for the studies of such state~\cite{PhysRevLett.87.192501}. 
Nevertheless, another excited state was identified as the $\alpha$-chain-like state~\cite{PhysRevC.84.054308,NEFF2004357,PhysRevC.91.024315}. 
Furthermore, the European group has suggested the possibility of four-$\alpha$ and 
six-$\alpha$ chain states~\cite{PhysRev.160.827,PhysRevLett.68.1295}.
From the theoretical side,
it has been discussed that giving angular momentum to the system or adding neutrons prevents the bending motion,
which is the main decay path, 
and stabilizes the 
linear-chain configurations~\cite{PhysRevC.64.014301,PhysRevC.74.067304,PhysRevC.83.021303,MARUHN20101,PhysRevLett.107.112501,PhysRevC.82.044301,PhysRevC.90.054307,PhysRevC.92.011303,PhysRevLett.115.022501}.

We can easily imagine that, if much longer linear $\alpha$-chain exists, it could be more stable than the other shape of aggregates with the same number of $\alpha$
particles due to the weaker Coulomb repulsion.  We examine the stability of 
longer linear chain state in terms of two kinds of resistances with a microscopic framework; the resistance against the complete break-up to $n\alpha$ particles and that against binary disintegrations to $\{m\alpha,(n-m)\alpha\}$, where $n$ and $m$ are integer numbers.  We have been enchanted by the simple beauty of $\alpha$-chain state. In other words, we strongly want to know one of the various facets of  $\alpha$-particle aggregation.

The aim of this report is to clarify the importance of the Pauli principle as well as the Coulomb interaction. 
For this purpose,
it could be a necessary condition 
to remove the ambiguity of the inter-nucleon force
by adopting the most reliable one 
suitable for the many $\alpha$-cluster aggregate.
We need the effective inter-nucleon force without any adjustable parameter;
having adjustable parameter
prevents us from comprehensively understanding the number dependence of the $\alpha$ clusters. 
The following conditions must be satisfied: (i) to guarantee overall saturation properties of nuclei, (ii) to naturally explain the binding energy and the size of the $\alpha$ particle, and (iii) to reproduce the phase shift of $\alpha$-$\alpha$ elastic scattering.  In short, we employ the effective interaction proposed by one of the authors (A.T.)
a quarter century ago, which is called F1 force~\cite{PhysRevC.49.1814},

\begin{equation}
\hat{V} = \sum_{i < j} V^{(2)}_{ij} + \sum_{i < j < k}  V^{(3)}_{ijk},
\end{equation}
where the two- and the three-body forces ($V^{(2)}_{ij}$ and $V^{(3)}_{ijk}$)
are written by the superposition of the Gaussian functions:
\begin{eqnarray}
&& V^{(2)}_{ij}  =  
\sum_{\l=1}^3 V^{(2)}_l \times \nonumber \\
&&
\{ (1-m^{(2)}_l) - m^{(2)}_l P^\sigma_{ij} P^\tau_{ij} \}
\exp[- (\vec r_i - \vec r_j )^2 / \beta_l^2],
\label{2body}
\end{eqnarray} 
\begin{eqnarray}
&& V^{(3)}_{ijk}  = 
\sum_{l=1}^3  V^{(3)}_l \times
\nonumber \\
 && 
\{(1-m_l^{(3)}) - m_l^{(3)}P^\sigma_{ij} P^\tau_{ij} \} \exp[- (\vec r_i - \vec r_j )^2 / \beta_l^2 ] \times
\nonumber \\
&&
\{(1-m_l^{(3)}) - m_l^{(3)}P^\sigma_{jk} P^\tau_{jk} \} \exp[- (\vec r_j - \vec r_k )^2 / \beta_l^2 ]. 
\end{eqnarray}
The force strengths, $V_l^{(2)}$ and $V_l^{(3)}$, and Majorana exchange 
strengths,  $m_l^{(2)}$  and $m_l^{(3)}$, are listed in Table~I.
The operator $P^\sigma_{ij}$ ($P^\tau_{ij}$) is for the exchange for spin state 
(isosopin state)
between $i$ and $j$ nucleons.
The introduction of the
three body force prevents the nuclear matter from a serious crush, and the finite-range effect
gives the optimal size of the $\alpha$ 
particles close to the experimental data.  The long range term expresses one pion exchange effect.

\begin{table} 
 \caption{ 
Parameter set for the interaction
(F1 force in Ref.~\cite{PhysRevC.49.1814}).}
  \begin{tabular}{cccccc} \hline \hline
 $l$  & $\beta_l$ (fm)  & $V^{(2)}_l$ (MeV) & $V^{(3)}_l$ (MeV) & $m^{(2)}_l$ &  $m^{(3)}_l$ \\ \hline
   1  &  2.5 & \ $-5.00$  & \ \ $-0.31$ & 0.75 & 0.00 \\  
   2  &  1.8 & $-43.51$& \ \ \ \ 7.73  & 0.462 & 0.00 \\  
   3  &  0.7 &\ \ $60.38$ &  219.0 & 0.522 & 1.909\\ \hline 
  \end{tabular}   \\ 
\label{Tohsaki-2}
\end{table}


Now we prepare the most intuitive model space to analyze the behavior of $\alpha$-chain state, which D.~M.~Brink has proposed a half-century ago~\cite{Brink};
a parameter space of $\bm{R}$ introduced to describe the geometrical configurations of $\alpha$ clusters, which is
called Brink-Bloch parameter space.
This is a fully microscopic framework, which takes full account of the Pauli principle. 
The wave function for $n\alpha$-clusters is written by

\begin{equation}
\Psi  = {\cal A}
\{
\phi({\bm R_1})
\phi({\bm R_2})
\cdots
\phi({\bm R_n})
\},
\label{total-wf}
\end{equation}
where ${\cal A}$ is the antisymmetrizer for all the nucleons, 
and the wave function of a single $\alpha$-particle wave function is defined by
\begin{equation}
\phi({\bm R_k}) =  \prod_{i,j=1,2}
\left(\frac{1}{\pi b^2} \right)^{\frac{3}{4}}
\exp \{- {1 \over 2b^2} \left({\bm r}^{ij}_k - {\bm R}_k \right)^{2} \} \chi^{ij}_k,
\label{Brink-wf}
\end{equation}
where $b$ is the nucleon width parameter, and $\chi^{ij}_k$ is a spin isospin wave function,
where superscripts $i$ and $j$ are labels for the spin and isospin.
The vectors $\{ {\bm r}^{ij}_k \}$ are the real coordinate for four nucleons in the $k$th $\alpha$-cluster, 
which shares the common Gaussian center ${\bm R_k}$.  
In this report, we adopt $b=1.415$~fm, which leads to the optimal binding energy of 
$\alpha$ particle; the quantity of 27.5~MeV 
is a little shallow but reasonable
compared with the experimental data (28.3~MeV).

The energy quantities are obtained by defining the norm and energy kernels,
$\langle \Psi | {\cal H} | \Psi \rangle$
and $\langle \Psi | \Psi \rangle$, 
where the Hamiltonian is written by
\begin{equation}
{\cal H} = -{\hbar^2 \over 2M}\sum_i \nabla_i^2 -T_g
+ \sum_{i < j} V^{(2)}_{ij}  + \sum_{i < j < k}  V^{(3)}_{ijk} + \sum_{i < j} V^{(c)}_{ij}.
\end{equation}
The first term is concerned with the kinetic energy, where $M$ is the nucleon mass.  
The second one is the center of mass energy to be removed.  
The following two terms describe the effective inter-nucleon force mentioned before.
The final term comes from the Coulomb energy.  
The adiabatic energy,
$E= \langle \Psi | {\cal H} | \Psi \rangle / \langle \Psi | \Psi \rangle$,
is a function of the configuration of the $\alpha$-clusters on the Brink-Bloch parameter space.  

Firstly, we prepare two sets of $\alpha$-cluster configurations, where equidistant 
$\alpha$ clusters are put on a straight line (called ``linear'') and on a circle (called ``annular''). 
We can obtain the total energy with respect to the distance between the nearest neighbor $\alpha$-clusters ($d_{in}$).  
In Fig.~1, we show the energies for the cases of $n=10$ and 15 $\alpha$-clusters ($n$: number of $\alpha$ clusters).
The horizontal axis is $d_{in}$ and the vertical axis is the binding energy per $n$, 
of which the origin is chosen for the $n\alpha$ break-up energy.  We can point out:
\begin{itemize}
\item[1.]
All the energy curves have the energy pocket, which guarantees the stability of $\alpha$-cluster chain state.  
In correspondence with the shallow energy pockets, there are hill-shaped barriers, where the Coulomb force plays an important role.  
\item[2.]
As for the case of 10 $\alpha$-clusters, the annular configuration is more stable than the linear configuration, on the other hand, the linear one is more stable than the annular one for the case with 15 $\alpha$-clusters. Therefore, the crossing point of the stability between two configurations exists between 10 and 15 $\alpha$-clusters.
\item[3.]
The positions of the energy pocket are almost independent of the two configuration sets. The points of the barrier have also the same feature; the property of 
the $\alpha$-$\alpha$ system in the free space is preserved.
\item[4.]
The long range parts behave like $\sim1/d_{in}$ due to the Coulomb force and show a gradual downward slope, which prevents the configurations from making a complete break-up of the $\alpha$-cluster chain.
\item[5.]
The energy pockets originate in synergetic effects of the attractive part of the inter-nucleon force and the Pauli principle of the total system, which works attractively. Inwardly, the Pauli principle works repulsively. This is well-known as the dual role of the Pauli principle~\cite{PTPS.E68.242}.
\end{itemize}

\begin{table} 
 \caption{ 
The physical quantities of $\alpha$-cluster chains. 
(a):	 the linear $\alpha$-cluster chain
(b):	 the annular $\alpha$-cluster chain.
}
 (a) \\
\begin{tabular}{c|cc|cc|c|c|c} \hline \hline
$n$ & $d_{min}$ & $E_{min}$ & $d_{max}$ & $E_{max}$ & $E_{max}-E_{min}$ & $E_{min}(l)$ & $N_{min}$  \\
     &     (fm)       &   (MeV)        &    (fm)         &     (MeV)        &         (MeV)       & $-E_{min}(a)$ & \\
\hline
5 & 3.20	& 8.455  & 7.07 &13.388 &4.933 &\ \ \ $2.511$   &  40 \\
10	& 3.25	&10.362 & 6.85 &15.463 &5.101	&\ \ \ $0.151$   &  180 \\
15	& 3.28	&11.301 & 6.73 &16.305 &5.005	& $-0.280$ &  420 \\
20	& 3.30	&11.917 & 6.65 &16.801 &4.884	& $-0.464$ &  760  \\
$\vdots$	& $\vdots$ & $\vdots$ & $\vdots$ & $\vdots$ & $\vdots$ & $\vdots$ &  $\vdots$ \\
50	& 3.35	&13.708 & 6.42 &18.054 &4.346	&$-0.756$ &  4900 \\
55	& 3.36	&13.884 & 6.40 &18.167 &4.282	&$-0.772$ &  5940 \\
60	& 3.36	&14.044 & 6.38 &18.268 &4.224	&$-0.785$ &  7080 \\
\hline \hline
\end{tabular}   

 (b) \\
\begin{tabular}{c|ccc|ccc|c} \hline \hline
$n$ & $d_{min}$ &$Rd_{min}$ & $E_{min}$	& $d_{max}$ & $Rd_{max}$ & $E_{max}$ & $E_{max}-E_{min}$ \\
      &     (fm)    &   (fm)        & (MeV)       &    (fm)       &   (fm)       &   (MeV)      &               \\
\hline
 5  & 2.84 &2.416 & 5.944	& 7.07	& 6.014 & 13.639 & 7.695 \\
10 & 3.21	&5.194 &10.210 & 6.77	& 10.955 & 15.834 & 5.624 \\
15 & 3.28	&7.888 & 11.580 & 6.63 & 15.945 & 16.718	& 5.138 \\
20 & 3.31	&10.580 & 12.381 & 6.54 & 20.903 & 17.235	 & 4.854 \\
$\vdots$	& $\vdots$  & $\vdots$ & $\vdots$ & $\vdots$ & $\vdots$ & $\vdots$ & $\vdots$ \\
50	& 3.38& 26.915 & 14.464 & 6.31 & 50.246 & 18.532 &4.068 \\
55	& 3.39	& 29.691 &14.656 & 6.29 & 55.090 & 18.648 &3.992 \\ 
60	& 3.39	&32.387 & 14.829 & 6.27 & 59.901 & 18.752 & 3.923 \\
\hline \hline
\end{tabular}    
\label{TableII}
\end{table}

\begin{table} 
 \caption{ 
The physical quantities of binary disintegrations of the linear $\alpha$-cluster chain.  
(a):	 the linear $\alpha$-cluster chain
(b):	 the annular $\alpha$-cluster chain.
}
 (a) \\
\begin{tabular}{c|cc|cc|c|c} \hline \hline
$n$ & $d_{min}$ &$E_{min}$ & $d_{max}$ & $E_{max}$ & $E_{max}-E_{min}$ & $E_{inf}$ \\
      &     (fm)     &   (MeV)   &    (fm)      &     (MeV)  &         (MeV)         & (MeV)  \\
\hline
 5 & 3.16	& $-0.005$ &6.99 & 6.171	& 6.176 & $-11.406$     \\
10 & 3.16	& $-0.024$ &6.86 & 5.978	& 6.002 & $-12.713$     \\ 
15	& 3.15	 & $-0.045$ & 6.82 & 5.922 & 5.967	& $-13.445$   \\
20	& 3.15& $-0.063$ &6.80 & 5.877	& 5.940 &$-13.960$    \\
$\vdots$	& $\vdots$ & $\vdots$ & $\vdots$ & $\vdots$ & $\vdots$ & $\vdots$  \\
50	& 3.14	 & $-0.121$ & 6.77	& 5.799 & 5.920	& $-15.546$   \\
55	& 3.14	& $-0.135$ & 6.77 & 5.788 & 5.923	& $-15.704$  \\ 
60	& 3.14	& $-0.135$ & 6.77 & 5.786 & 5.921	& $-15.861$  \\ 
\hline \hline
\end{tabular}   

 (b) \\
\begin{tabular}{c|cc|cc|c|c} \hline \hline
$n$ & $d_{min}$ & $E_{min}$ & $d_{max}$ & $E_{max}$ & $E_{max}-E_{min}$ & $E_{inf}$ \\
     &     (fm)     &   (MeV)   &    (fm)      &     (MeV)  &         (MeV)         & (MeV) \\
\hline
 5 & 3.25	&$-0.007$ &6.80& 5.823 &5.830	&$-12.838$ \\
10	& 3.30	&$-0.008$ &6.35& 4.791 &4.799	&$-19.066$\\
15	& 3.32	&$-0.005$ &6.15& 4.267 &4.272	&$-25.042$\\
20	& 3.34	&$-0.005$ &6.02& 3.901 &3.906	&$-31.106$\\
$\vdots$	& $\vdots$ & $\vdots$ & $\vdots$ & $\vdots$ & $\vdots$ & $\vdots$ \\
50	& 3.40	&$-0.008$ &5.68& 2.852 &2.860	&$-66.735$\\
55	& 3.41	&$-0.006$ &5.65& 2.765 &2.771	&$-72.605$\\
60	& 3.42	&$-0.009$ &5.62& 2.665 &2.674	&$-78.504$\\
\hline \hline
\end{tabular}    
\label{TableIII}
\end{table}

The general traits of the depth of the pocket and the heights of the barrier (per $n$)
are shown in Fig.~2, where $n$ is increased up to 60.
The general trends can be seen here,
but to see more details, we list  in Table~II the energies
of the first four cases and the last three cases when $n$ is increased to 60 with the step of 5. Table II~(a) corresponds to the case of linear $\alpha$-cluster chain. 
The first column ($n$) is the number of $\alpha$-clusters, the second column ($d_{min}$)
means the value of  $d_{in}$ at the energy pocket, the third column ($E_{min}$)
is the minimum energy there, the fourth column ($d_{max}$)
is the $d_{in}$ value at the hill-top, and the fifth column ($E_{max}$)
is the barrier energy there. The final two columns are devoted to the depth of the pocket measured from the barrier and the deviation of the energy of the linear chain from that of the annular one. Table II~(b) stands for the case of the annular chain of $\alpha$-clusters. Two columns are added, which are those of the annular radii;
$Rd_{mim}$
$(Rd_{max})$
 corresponds to the radius of the energy pocket (hill-top).   Beautiful systematics are shown as follows:
\begin{itemize}
\item[1.]
Both linear and annular configurations have the energy pocket up to 60 $\alpha$-clusters. It could be imagined that such tendency maintains for vast numbers of $\alpha$ clusters. 
\item[2.]
The difference between the barrier height and energy pocket per $n$ ($E_{max}-E_{min}$ )
are converged to 4.2~MeV and 3.9~MeV for the linear (Table~II~(a)) and annular (Table~II~(b)) configurations, respectively. This deviation comes from the geometrical difference of the configurations, which affects the contributions of the Coulomb energy and the Pauli principle.
\item[3.]
The positions of the energy pocket ($d_{min}$) are almost the same in large $n$ cases.  The fact that the nearest neighbor distance of $\alpha$-clusters ($d_{in}$) is about 3.3~fm means again the conservation of the property of the individual $\alpha$-$\alpha$ cluster structure in the free space. 
\item[4.]
The four curves in Fig.~2 behave like $\sum_{k=1}^n (1/k) \sim \log(n) +\gamma$, where $\gamma$ is Euler constant,
0.57721$\cdots$. In other words, as far as the long $\alpha$-cluster chain exists, the energy with respect to the number of the $\alpha$ clusters diverges as logarithmic behavior. The reason comes from the fact that the other energies except for the Coulomb energy, that is, the kinetic and the interaction energy originating in the nuclear force, are saturated as the number of $\alpha$ clusters increases.  
\item[5.]
The linear configuration becomes more stable than annular one beyond $n=12$.  The reason also comes from the fact that the reduction of the Coulomb energy works well for the linear configuration compared with the annular one.

\end{itemize}

We can point out that
if the $\alpha$-cluster chain is on various lines, such as linear, annular, and spiral lines, the linear configuration gives the lowest energy,
as the number of the $\alpha$ clusters increases. 
The column $E_{min}(l)-E_{min}(a)$ in Table~II~(a) shows the difference between the minimum energies of linear and annular chain states.
Nevertheless, many kinds of long $\alpha$-cluster chains can exist stably. We cannot find the limiting numbers of $\alpha$-clusters for a complete break-up.

\begin{figure}[t]
	\centering
	\includegraphics[width=6.5cm]{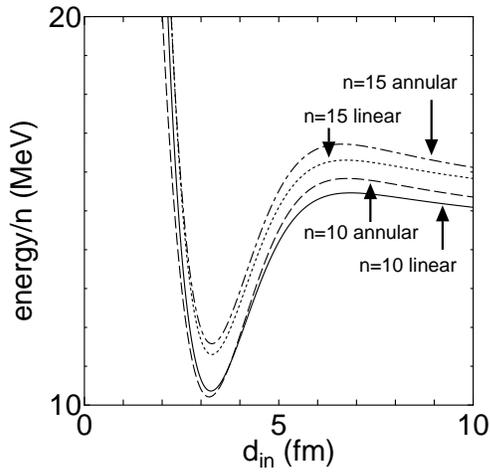} 
	\caption{
The energy per number of $\alpha$ clusters ($n$)
versus the nearest neighbor distance between equally separated $\alpha$-clusters ($d_{in}$).  
The cases of $n=10, 15$  are shown.
     }
\label{fig1}
\end{figure}

\begin{figure}[t]
	\centering
	\includegraphics[width=6.5cm]{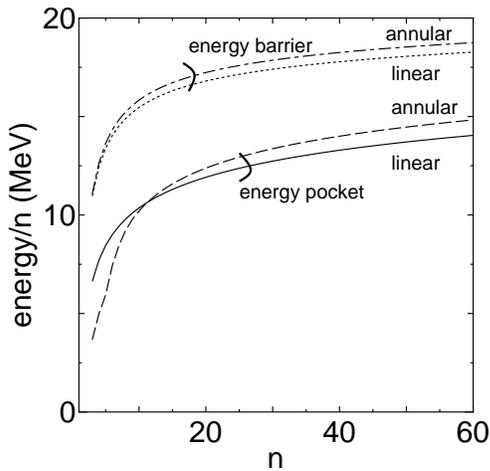} 
	\caption{
The systematics of the energy pocket and energy barrier as functions of number of equally separated $\alpha$ clusters ($n$)
for the linear and annular configurations.
     }
\label{fig2}
\end{figure}

\begin{figure}[t]
	\centering
	\includegraphics[width=6.5cm]{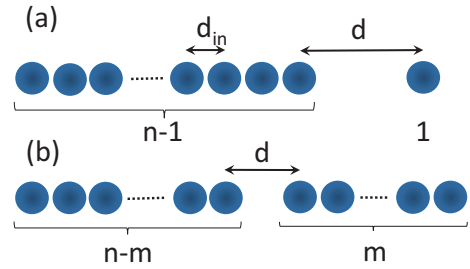} 
	\caption{
Schematic figure for the linear configuration of $n\alpha$ clusters.
(a): $\alpha$ decay of one $\alpha$ cluster.
(b): fission into $(n-m)\alpha$ cluster and $m\alpha$ cluster systems.
The relative distance between the two systems is $d$,
whereas the $\alpha$-$\alpha$ distance is $d_{in}$.
     }
\label{fig3}
\end{figure}

\begin{figure}[t]
\includegraphics[width=6.5cm]{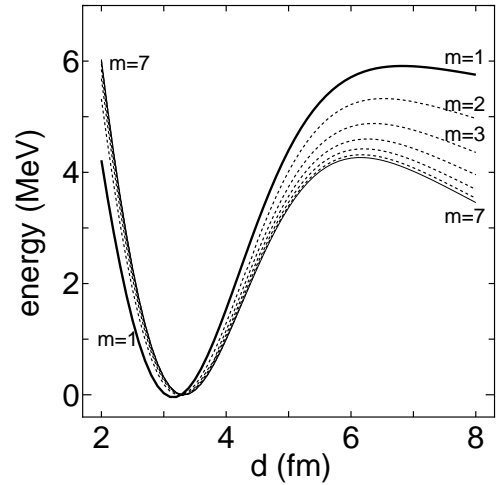} 
	\caption{
The energy curves versus $d$ (fm), the nearest neighbor distance between $\alpha$-cluster chains in the different blocks. The 15 $\alpha$ clusters are 
separated to two blocks with $m$ and $15-m$ $\alpha$ clusters.
     }
\label{fig4}
\end{figure}

\begin{figure}[t]
	\centering
	\includegraphics[width=6.5cm]{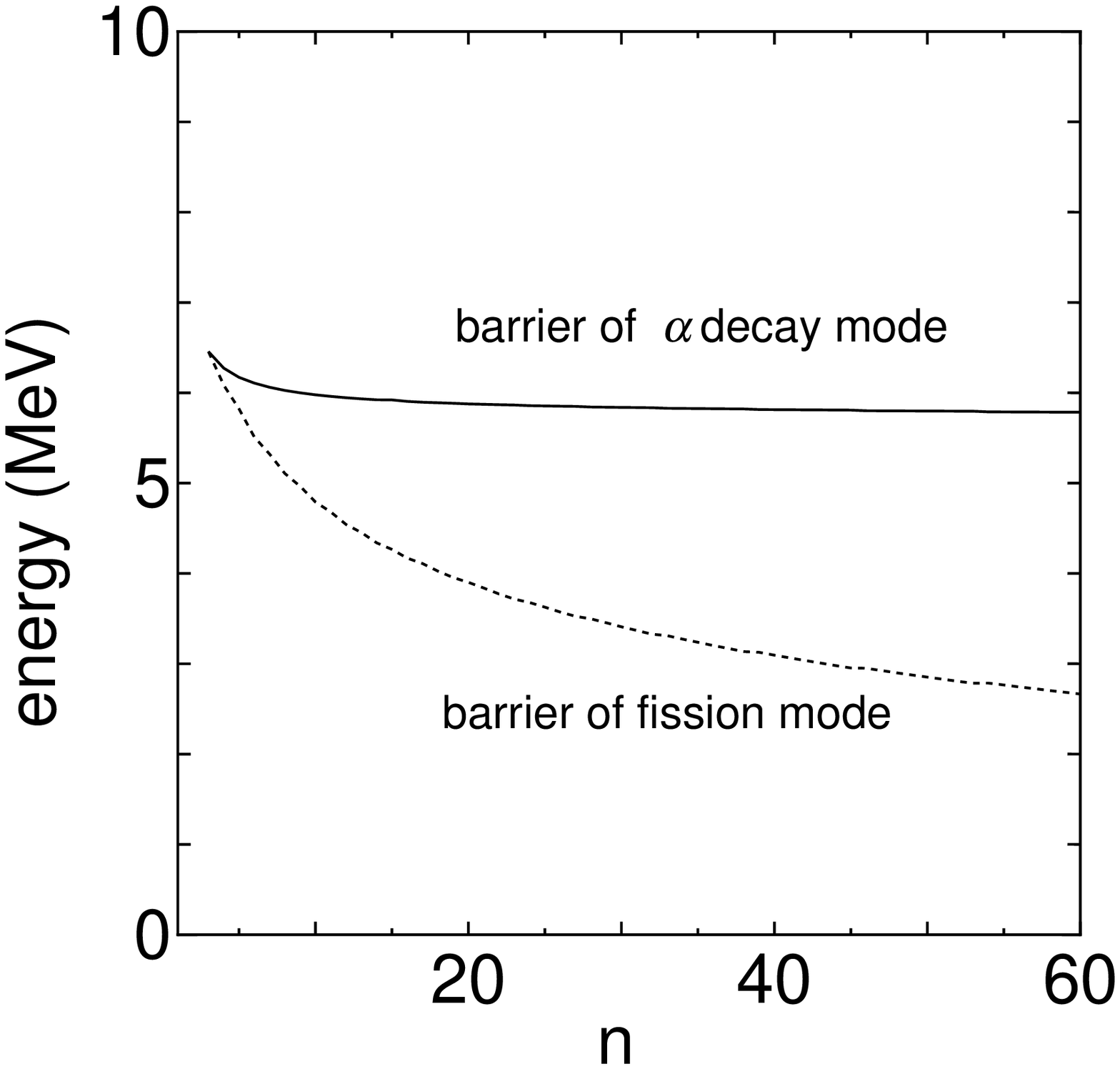} 
	\caption{
The barrier height of  $\alpha$-decay mode and the fission mode.
     }
\label{fig5}
\end{figure}

Secondly, let us scrutinize the stability of the linear $\alpha$-cluster chain in terms of binary disintegration.  
In Fig.~3, a schematic figure is shown.  
The $n$ $\alpha$-clusters are divided into two blocks, that is, one block with $m$ $\alpha$-clusters and the other with 
$(n-m)$ $\alpha$-clusters.  The distance, $d$, between nearest neighbor $\alpha$-clusters in the different blocks is a variable parameter to see the stability of the disintegration. The case with $m=1$ is regarded as the mode corresponding to $\alpha$-decay process, and the case with $m \sim n/2$ ($m = n/2$ for even $n$, $m=(n-1)/2$ for odd $n$) describes the symmetric fission. The inner distance of two $\alpha$-clusters, $d_{in}$, is taken as that leads to the minimum energy of linear $n\alpha$-clusters. 
We show the energy curves for 15 $\alpha$-clusters case in Fig.~4.  
The seven curves corresponding to different $m$ are depicted versus the distance $d$. 
Unlike Fig.~1, Fig.~2, and Table~II, where we discussed the total energy per $n$, from here we employ the total energy (not per $n$).
The energy is measured from the pocket of the equidistant linear chain, 
and at $d \to \infty$,
the energy reaches an individual one depending on $m$.
We can learn that all the energies for the binary disintegration have the energy minimum with the $d$ value, 
which is close to the optimal $d_{in}$ value of the equidistant 15 $ \alpha$ clusters on a straight line. 
Furthermore, we can verify that the minimum energies come to the energy of the pocket of the equidistant linear chain. 
This means that 15 $\alpha$-cluster chain is stable against any binary disintegration.  
The maximum energy around $d \sim 6$~fm is continuously changed depending on the number of $m$, and the $\alpha$-decay mode seems 
to be more stable than the fission one, as the height decreases with increasing $m$.  
In detail, the optimal $d$ value for the $\alpha$-decay mode ($m=1$) slightly deviates inwardly from the optimal $d_{in}$ value of the equidistant model. 
Therefore, the equidistant model should be a little modified, because the edge effect takes place,
which stabilizes the chain-state. 

Here we discuss the quanta of the harmonic oscillator, $N$. 
The lowest allowed quanta are the ones obtained at the zero distance limit between the $\alpha$ clusters ($d_{in} \to 0$),
shown in the column $N_{min}$ of Table~II~(a). 
It is seen that the $N_{min}$ value decreases, when the linear chain state is separated to two blocks;
 $N_{min}$ is 420 for $n=15$, and it decreases to $40+180=220$ after separated into 5 $\alpha$'s and 10 $\alpha$'s ($m=5$).
 In the $n=15$ case, this decrease of  $N_{min}$ is smallest (56) at $m=1$ and largest (224) at $m=7$.
Indeed, the decrease of $N_{min}$ reaches the largest value in the case of symmetric fission,
which is related to the decrease of the peak energy around $d \sim 6$~fm in Fig.~4 with increasing $m$.

Focusing the $\alpha$-decay mode ($m=1$) and the symmetric fission mode ($m=n/2$), we show in Fig.~5
the barrier height, which is the height of the local peak around $d \sim 6$~fm, 
with respect to the number of $\alpha$ clusters;
the bold line stands for the $\alpha$-decay mode and the dotted one is for the fission mode.  
The $\alpha$-decay mode reveals almost the same height through all the number of $\alpha$ clusters,  
but the fission mode looks like an exponential decrease.
Therefore, the fission mode is more fragile than $\alpha$-decay mode,
and adding external forces may strongly affect the stability of $\alpha$-cluster chain.

The physical quantities for the binary disintegrations 
are summarized in Table~III, 
(a):	 the linear $\alpha$-cluster chain
(b):	 the annular $\alpha$-cluster chain.
Here,
$d_{min}$ is the $d$ value for the energy pocket, $E_{min}$ is the minimum energy there,
$d_{max}$ is the $d$ value for the local peak,
and $E_{max}$ is the perk energy there.
The final column ($E_{inf}$) is the total energy when two blocks are separated into
two independent systems as $d \to \infty$.

Let us summarize the energy properties of $\alpha$-chain states. As the first stage, we examined the adiabatic energy curves with respect to two kinds of distances on the Brink-Bloch parameter space within the microscopic framework. 
The Pauli principle and the exact treatment of the Coulomb force are certainly important for understanding the energy properties, which have been completely taken into account.
Furthermore, the most appropriate effective inter-nucleon force, which can reproduce overall saturation property of nuclei and the experimental data of the $\alpha$-$\alpha$ elastic scattering, has to be adopted, and here we introduced the F1 force.

As for the $\alpha$-cluster chain with equidistant 
distribution, the nearest neighbor distance is a suitable measure for the instant break-up of the chain. 
In this respect, we pointed out that the $\alpha$-cluster chain is always stable against the complete break-up. 
This is because each term of the Hamiltonian,  except for the Coulomb energy,
is saturated with increasing number of the $\alpha$-clusters if the expectation value is divided by the number.
Nevertheless, the effect of the Coulomb energy is too weak to break the chain,
and the 
optimal $\alpha$-$\alpha$ distance stays the value almost the same as the free $\alpha$-$\alpha$ system. 
As for the disintegration into two blocks of the linear chain, we examined the stability with respect to 
the distance between the edges of the two blocks. 
We can learn that
once $\alpha$-cluster linear chain is formed, it is  surrounded by a long-tail wall owing to the Coulomb force between two blocks. The Pauli principle works attractively inward in the barrier.  These effects stabilize the linear chain states and prevent the break-up.

In this report, we did not mention the bending motion of the chain, which is the deviation
from the one-dimensional configuration,
but we already verified the stability of the linear chain of $\alpha$-clusters against it.  In view of the reduction of the Coulomb energy, the linear chain always remains unchanged, and the reduction of the Coulomb energy
is quite important as the number of $\alpha$ clusters increases.  
Unfortunately, we do not know any appropriate effective inter-nucleon force to study the overall trends of $\alpha$-cluster aggregate except for F1 force. We have estimated the energy property of the $\alpha$-chain state using the Brink-Boeker force~\cite{BRINK19671}, 
which also do not contain adjustable parameters. Quantitative energy obtained is different from each other, but the tendency of 
the chain state is not so changed comparing with that of the F1 force. We will report detailed results on the dependence of the effective inter-nucleon force elsewhere. 

We imagine that various kinds of $\alpha$-cluster chain float in the space, above all the linear chain, possibly exists.  The chain could be plausibly disintegrated by the fission mode when an external force acts on it.

\begin{acknowledgments}
We are indebted to Prof.~N. Tajima who has suggested the instability for the fission mode.
The numerical calculation has been performed using the computer facility of 
Yukawa Institute for Theoretical Physics,
Kyoto University. This work was supported by JSPS KAKENHI Grant Number 17K05440.
\end{acknowledgments}

\bibliographystyle{apsrev4-1}
\bibliography{biblio_ni.bib}

\end{document}